\documentclass{elsart}
\usepackage{epsfig,amssymb}
\usepackage{rotating}

\begin{document}

\begin{frontmatter}



\title{A new path toward gravity experiments with anti-hydrogen}


\author{P. Perez,} 
\ead{patrice.perez@cea.fr} 
\author{A. Rosowsky}
\ead{andre.rosowsky@cern.ch}
\address{DSM/Dapnia/SPP, CEA/Saclay, F-91191 GIF-SUR-YVETTE}

\begin{abstract}
We propose to use a 13 KeV antiproton beam passing through a dense
cloud of positronium (Ps) atoms 
to produce an $\mathrm{\overline{H}^+}$ ``beam''. 
These ions can be slowed down and 
captured by a trap. The process involves two reactions with 
large cross sections under the same experimental conditions. 
These reactions are the interaction of $\mathrm{\overline{p}}$ with
$\mathrm{P_{S}}$ to produce $\mathrm{\overline{H}}$ and 
the $e^{+}$ capture by $\mathrm{\overline{H}}$ reacting on $\mathrm{P_{S}}$ 
to produce $\mathrm{\overline{H}^+}$. Once 
decelerated with an electrostatic field and 
captured in a trap the $\mathrm{\overline{H}^+}$ ions could be cooled and the 
$e^{+}$ removed with a laser to perform a measurement of the 
gravitational acceleration of neutral antimatter in the gravity field 
of the Earth. 
\end{abstract}

\end{frontmatter}

\section{Introduction}
The measurement of the gravitational mass of antiparticles motivates 
physicists since over three decades. It is interesting to note that 
general relativity would not contradict antiparticles to ``fall up" in 
the gravity field of the earth~\cite{carter}. Experiments to test such ideas
have been proposed on positrons and antiprotons~\cite{antigravprop} but 
never succeeded. Several experiments at CERN are now producing neutral 
antimatter in the form of antihydrogen or antiprotonic helium
atoms~\cite{cernantih,gabrielse} for CPT tests. The trapping of
neutral antihydrogen 
atoms is the next step for these experiments. Some proposals to measure the 
fall of these atoms have been presented~\cite{cesar}.  
The possibility to measure the free fall of positronium has also been 
studied~\cite{mills_Psfall}.

\par
Recently it has been proposed~\cite{walz,waltz2-3} to measure the
gravity acceleration of antimatter 
using the $\mathrm{\overline{H}^+}$ ion. This ion has the advantage that it 
can be cooled down to 13 $\mu$K, a temperature suitable for a gravity 
experiment. 
The authors do not investigate in detail the  $\mathrm{\overline{H}^+}$ 
production scheme but suggest the interaction of $\mathrm{P_{S}}$ 
Rydberg atoms with  $\mathrm{\overline{H}}$
atoms~\cite{charlton1990,psstar}. 
In this article 
we concentrate on the  $\mathrm{\overline{H}^+}$ production scheme and propose
a ``beam to cloud'' experimental configuration instead of a trapped system.
This configuration allows the isolation of the
$\mathrm{\overline{H}^+}$ from the production region.

\par
The positronium cloud hereafter referred to as ``the target'' is not
contained: it is made of positronium atoms emitted from an Aluminium
crystal surface which is bombarded by a flux of positrons. 
The atoms are emitted in a direction normal to the surface with a few 
mrd angular spread~\cite{Ps_emission}. 
Their energy is $\sim 2$ eV and their speed is $\sim$1 mm/ns. 
The density of the Ps target is proportional to the positron flux.
As explained further, an amount of order $10^{11}$
positrons is needed in order to get the required density for the Ps
target. 
However, the short Ps lifetime requires this amount to be delivered
in a few nanoseconds while the highest foreseen rates of slow 
positrons are at best $10^{11} \ \mathrm{s^{-1}}$. 
The positrons have then to be accumulated in the crystal vicinity and 
accelerated toward
the crystal. 
To counteract the effects of space charge, the positrons are
accumulated into a 
small neutral $e^{-}e^{+}$ ``plasma'' for a short time.
The plasma size is $\sim 3 \mathrm{mm^2 \times 1 cm}$.
Then an electrostatic field accelerates the positrons toward an
Aluminium crystal where they are converted into Positronium atoms. 

\par
A 13 keV antiproton beam is guided parallel to the crystal at a distance of 
$ \sim 150 \ \mu$m. 
The target length is crossed in 6 ns by the 13 keV antiprotons.

\par
\begin{figure}
%\vspace{-1cm}
\begin{center}
%\begin{picture}(100,90)
%\put(10,10){\epsfxsize85mm\epsfbox{synoptic.eps}}
\includegraphics*[width=140mm]{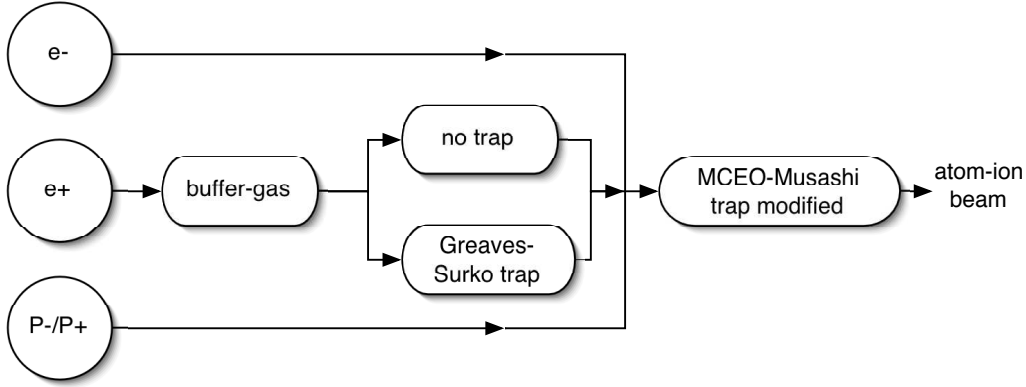}
%\end{picture}
\end{center}
\vspace*{-0.3cm} 
\caption{\protect\footnotesize 
 Atoms/ions production scheme.}
\label{fig:synoptic}
\end{figure}

The flux of positronium atoms is maintained as long as the flux of positrons
toward the crystal is kept. 
It is the flux of positronium atoms which constitutes the ``target''. 
The overall layout for such a gravity experiment is presented
in figure~\ref{fig:synoptic}.

\par
Aside from the gravity experiment, 
the technique to accumulate positrons in the plasma volume described in this 
article could prove useful as a first step toward the realization of a
positronium~\cite{platzman_mills} BEC, a 511 KeV laser~\cite{mills-laser} and 
the observation of the antimatter molecule Ps$_{2}$.
When Greaves-Surko traps~\cite{traps,surko-trap,traplim} are available with a 
capability to hold $10^{12}$ positrons and provided that the neutral
plasma can be held during the 
time required to empty the trap, one could produce a density of positronium of 
$\sim 0.3 \ 10^{14}$ cm$^{-3}$. An experiment to observe 
the stimulated annihilation process would then become feasible.

\par 
If traps with such capability are not available, an alternative path toward a
stimulated annihilation observation could be to use the neutral plasma 
in a MCEO trap~\cite{musashi}.
The positronium atoms produced by the 3 body reaction are in  Rydberg states.
It takes a few $\mu$s before they reach the ground state. 
There, the magnetic field couples the triplet state to the singlet
state, which has a lifetime of 0.125 ns.
Therefore the life time of the positronium produced inside the plasma 
is dominated by the decay time towards the ground state.
Using an infrared laser onto the $\mathrm{P_{s}^{*}}$ gas one can 
re-ionize some atoms while preventing the decay into states with a low
level of excitation. Hence the life time can be increased and, in principle, 
longer trains of positrons could be filled into the plasma volume.

\par
This configuration is also well suited for spectroscopy experiments where atoms 
containing a positron, or a positronium atom~\cite{Ps-chemistry,psstar} are 
produced by the interaction of atoms/ions with the target. 
The incoming antiproton beam is then replaced by an ion or atom beam.

\par
The path described in this article to produce an antimatter ion beam
is the following:

\begin{itemize}
\item the accumulation of positrons in a neutral $e^{-}e^{+}$ plasma,
\item the separation of $e^{-}$ and $e^{+}$ by an electrostatic field and 
the interaction of the $e^{+}$ with an Aluminium crystal to produce a 
positronium cloud referred as the target,
\item the interaction of 13 keV (anti)protons with the target
to produce the atoms and the ions as a beam.
\end{itemize}

\par
The beam to cloud configuration presented in this article is made of
several devices which were developped separately for various applications:
\begin{itemize}
\item the 10 MeV $e^-$ beam on a thin foil to produce an intense $e^+$ source,
\item the buffer gas and Greaves-Surko trap,
\item the MCEO trap,
\item the Charge Focusing Aluminium Converter (CFAC).
\end{itemize}

The present study takes the parameters of these devices as 
they appear in the litterature and shows that with little modifications,
these devices can be assembled to produce an $\overline{H}^+$ ion beam.

The steps involved in the target production are discussed in the next section.
The atom and ion production rates are presented in section~\ref{antihprod}.
Several technical features are discussed in section~\ref{discuss}.

\section{The positronium target}
\label{pstarget}

The positronium target is obtained by accelerating positrons from an 
$e^{-}e^{+}$ plasma toward an Aluminium crystal. 
The positrons hit the crystal with a kinetic energy above 40 eV to
avoid elastic and specular reflection~\cite{Ps_specular}.

\par
In the following we describe a way to produce the neutral plasma and 
the subsequent extraction and
focalisation of the positrons from the plasma onto the Al crystal.

We foresee two modes of operation for the creation of the neutral plasma:

\begin{itemize}
\item the slow loading mode where the positrons are extracted
 continuously from the buffer gas section of the Greaves-Surko trap 
 which cools them to room temperature ($\sim$ 25 meV). The loading time is 
 a few seconds.
\item the fast loading mode where the positrons are stored and cooled
 to 2 meV in a 
 Greaves-Surko trap~\cite{traplim} and extracted in $\sim$ 10 $\mu$s. 
 This fast extraction heats the positron beam. This mode of operation requires 
 less confinement time for the neutral plasma and may allow higher plasma densities. 
\end{itemize}

The electrons are provided by a buffer gas or magnetic trap depending
on the desired temperature. Since electrons do not annihilate on the
container walls and are easy to produce, the final neutral plasma 
temperature is tuned by setting the electron beam temperature.

We assume that the positrons are produced continuously through the 
interaction of and intense beam of 10 MeV electrons with a thin tungsten 
foil~\cite{perez-roso} with a rate, after solid Neon moderation, 
of $10^{11} \ \mathrm{s^{-1}}$.
The positrons are cooled to room temperature ($\sim 25$ meV) in a
buffer gas. 
In the slow loading mode the positrons are taken at the exit of the 
buffer gas section while in the fast loading mode they are first
accumulated in a Greaves-Surko trap~\cite{traplim,traps,surko-trap} 
where they are
stored and cooled to $\sim 2$ meV. The low temperature of the positrons 
is a feature of the Greaves-Surko trap: the strong magnetic field is 
produced by a 
supraconducting magnet and the positrons are in thermal equilibrium
with the magnet container at a temperature of 2 meV. Such a low temperature
enhances the 3 body reaction which absorbs the positrons if the neutral plasma were
to be stored in this trap. 

\par
The preferred device for accumulating the $e^+$ in a neutral plasma is 
the MCEO trap~\cite{musashi} which does not require a very low temperature
such as the one reached in a Greaves-Surko trap.
Therefore the heating of the positrons can be accomodated
and is beneficial to accumulate enough positrons in the neutral plasma.

The positrons are guided by a 100 Gauss magnetic field, from the 
Greaves-Surko trap (or the buffer gas), to the neutral plasma.
The injection pipe is shielded to minimize its leaking field.

\begin{figure}
%\vspace{-1cm}
\begin{center}
%\begin{picture}(100,90)
%\put(10,10){\epsfxsize85mm\epsfbox{cusp.eps}}
%\includegraphics*[width=140mm]{cusp.eps}
\includegraphics*[width=90mm]{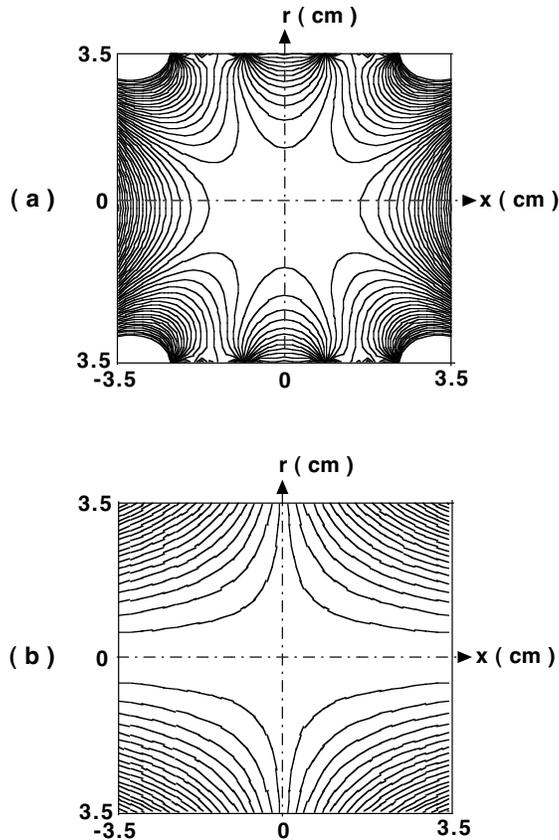}
%\end{picture}
\end{center}
\vspace*{-0.3cm} 
\caption{\protect\footnotesize 
 The MCEO trap electric potential (a) and  magnetic flux (b)
 set for antihydrogen synthesis~\cite{musashi}.}
\label{fig:cusp}
\end{figure}

The original MCEO trap is a magnetic cusp with an octupolar electric field.
It is made of several pairs of coils of 10 cm radius arranged symmetrically with 
respect to the center of the trap. 
The most central coils are located at 4 cm from the center. The other
coils are separated from each other by 5 cm.
The currents in each pair of symmetric coils are of opposite sign. 
The current amplitude can be varied for each pair.

It is designed to mix both 
positrons and antiprotons. It is to be implemented in the
ASACUSA beam line at CERN under the name of Musashi trap.
In our application this trap is meant to mix 
positrons with electrons. Therefore a scaled down version has to be
designed.

The scaled down version of the MCEO trap has the following parameters:
the magnetic coils radius is 1.5 cm, the most central coils are at 1.2 cm from the center,
the next coils are separated by 1.5 cm, the total current for each set of coils 
is 4 k Ampere-turns, the octupolar electrostatic field is set by a central electrode
at 70 V and a length of 1.0 cm.
The key feature of this trap
is to have a null field region in its center (figure~\ref{fig:cusp}). Once a charged plasma is
loaded, it acts as a potential well for the opposite charge. We use
electrons to create the well for the positrons. The neutral plasma
will be located at the center of the well.
The $e^{-}$ and $e^{+}$ beams are injected along the x axis.
The octupolar electrostatic field of the MCEO trap is simulated
using an analytical form~\cite{octapole}.

A simple trap made of two coils acting as mirrors~\cite{bottle} is
also considered but the injection
of the positrons is more difficult.

\begin{figure}
\vspace{-3cm}
\begin{center}
%\begin{picture}(100,90)
%\put(10,10){\epsfxsize85mm\epsfbox{fig-soudure1.eps}}
\includegraphics*[width=100mm]{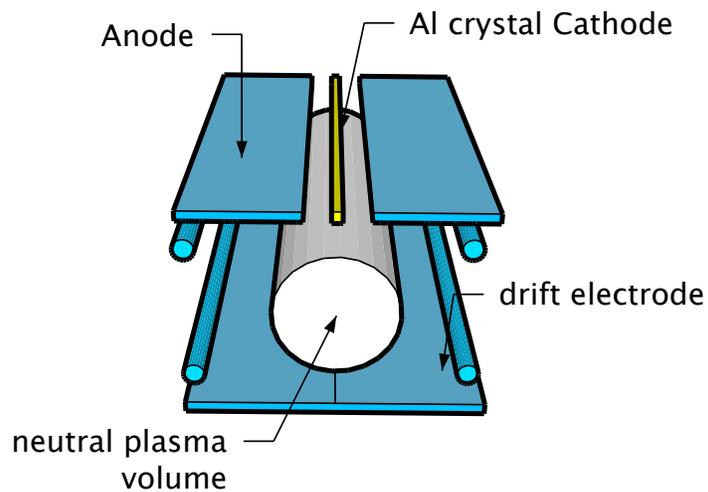}
%\end{picture}
\end{center}
\vspace*{-3.0cm} 
\caption{\protect\footnotesize 
 The CFAC electrodes geometry.}
\label{fig:msgc1a}
\end{figure}

In order to extract the positrons from the neutral plasma and
accelerate them toward the crystal, an electrostatic field
is used. The electrode configuration is shown in figure \ref{fig:msgc1a}.

The electrodes are arranged in two parallel planes. The cathode is
made of Al cristal.
Its surface is 100 $\mu$m x 1 cm. The neutral plasma is located in the 
volume between the two planes. This configuration will be referred as 
Charge Focusing Aluminium Converter or CFAC.

This configuration is inspired by a technology
developped for high energy physics detectors called the 
Micro Strip Gas Chamber~\cite{MSGCform}, or MSGC.
The distances between electrodes and field magnitude are taken from 
this development.

\begin{figure}
%\vspace{-1cm}
\begin{center}
%\begin{picture}(100,90)
%\put(10,10){\epsfxsize85mm\epsfbox{fig-soudure1.eps}}
\includegraphics*[width=90mm]{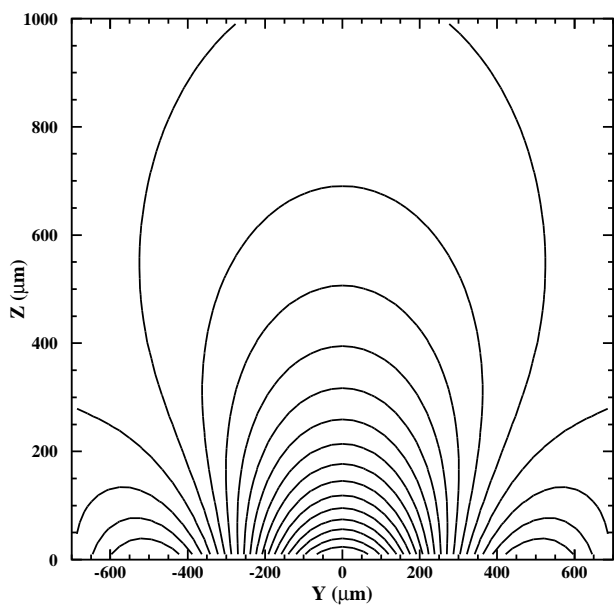}
%\end{picture}
\end{center}
\vspace*{-0.3cm} 
\caption{\protect\footnotesize 
 The CFAC electrostatic potential: (0,0) is the center of the crystal surface.}
\label{fig:msgcpot}
\end{figure}

In the CFAC, the electrostatic field of the MSGC is modified by introducing 4 wires
to smoothe the side field as shown in figure \ref{fig:msgcpot}.

The CFAC was put in the center 
of the trap by the simulation described below in order to get an estimation of the 
positronium density and time distribution.
The insertion of metallic conductors in the 
center of the MCEO trap is unrealistic. An integrated design is 
proposed in the discussion section. 

A fortran program was written to simulate:
\begin{itemize}
\item the stability of the neutral plasma during the 10 $\mu$s loading time,
\item the acceleration by the CFAC electrostatic field and the time 
distribution of the positrons hitting the Al crystal.
\end{itemize}

\subsection{The target simulation}
In the MCEO trap the positrons are injected along its axis, 
parallel to the magnetic field which guides them.

\subsubsection{The CFAC electrostatic field}
The CFAC electrostatic field is computed using analytical 
formulae~\cite{MSGCform}. 
The gap of the CFAC,
i.e. distance between the drift electrode and the cathode planes is 1.5 cm. The
cathode plane is located at 1 cm from the x axis and parallel to it.
The drift electrode size is 1 cm x 7 mm. The cathode, made of the
Al(111) crystal has a 100 $\mu$m width. The distance between the edge
of the cathode and the next electrode in the cathode plane is 200 $\mu$m.
The drift electrode plane being at 1600 V, the cathode is grounded
and the bias voltage is 1550 V. The plasma temperature is 30 meV. 

During the loading phase the CFAC field is set ``off''.
The acceleration phase starts when the CFAC electrostatic field is set
``on'' while the octuplole electrostatic field and the cusp magnetic field are set ''off''.

\begin{figure}
%\vspace{-1cm}
\begin{center}
%\begin{picture}(100,90)
%\put(10,10){\epsfxsize85mm\epsfbox{fig-soudure1.eps}}
\includegraphics*[width=85mm]{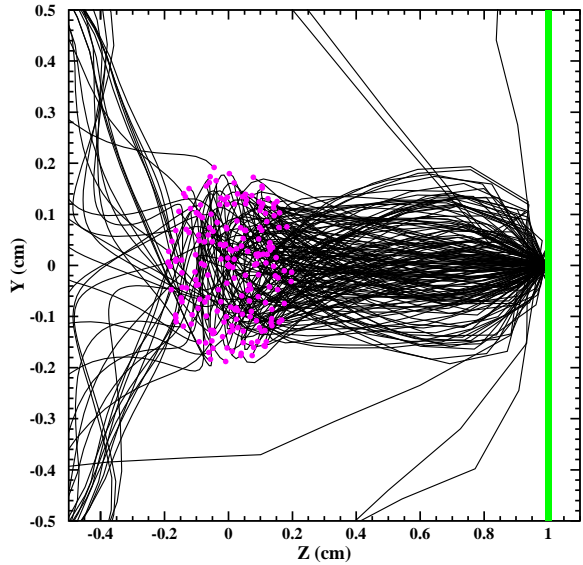}
%\end{picture}
\end{center}
\vspace*{-0.3cm} 
\caption{\protect\footnotesize 
Positron tracks from the plasma to the CFAC in cm.}
\label{fig:msgc_xyz}
\end{figure}

\begin{figure}
%\vspace{-1cm}
\begin{center}
%\begin{picture}(100,90)
%\put(10,10){\epsfxsize85mm\epsfbox{fig-soudure1.eps}}
\includegraphics*[width=85mm]{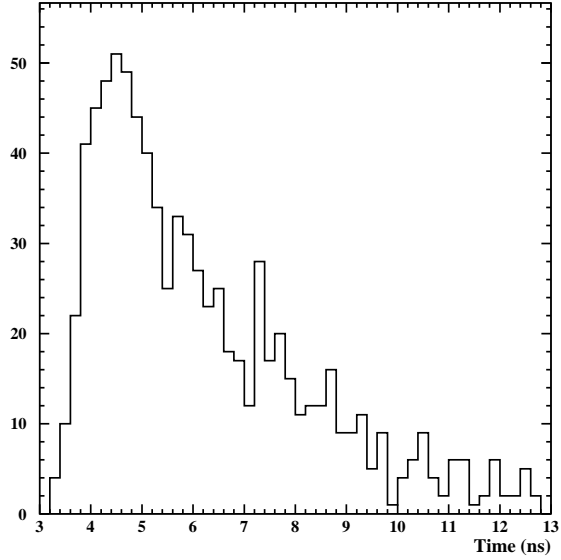}
%\end{picture}
\end{center}
\vspace*{-0.3cm} 
\caption{\protect\footnotesize 
Arrival time distribution of the positrons on the crystal in nanoseconds.}
\label{fig:msgc_tim}
\end{figure}

The positron capture and the emission of positronium atoms is a fast 
process, the delay of which can be neglected at the nanosecond 
scale~\cite{lynn-review}. 
Therefore the time distribution of the positrons hitting
the crystal is also the time distribution of the positronium atoms 
exiting the crystal surface (figures \ref{fig:msgc_xyz},\ref{fig:msgc_tim}). 

The positronium atoms exiting the crystal are in the ground state. 
The magnetic field couples the singlet and triplet states~\cite{Ps_quench},
reducing the Positronium life time to a minimum of 0.125 ns.

In an experimental test with a MCEO trap and
with electrons the radial field on the central electrode surface
was 360 Gauss~\cite{leap03}.
This trap is designed to hold both positrons and
antiprotons. When antiprotons are used the magnetic field is much stronger but
in the beam to cloud configuration the antiprotons cross the trap axis 
and are not kept. Hence with a weak field ($\sim$ 3.5 Gauss) we take as effective
lifetime 0.200 ns.
This life time corresponds to a travel distance of 200 $\mu$m.
The target volume is defined as a box parallel to the crystal,
which starts at a distance of 100 $\mu$m. The target is 100 $\mu$m
large, 1 cm long and 100 $\mu$m thick,
which gives a volume of $10^{-4} \ \mathrm{cm^3}$. 
This target volume is centered at a distance of 150 $\mu$m from the 
crystal surface.
%From the crystal width of 100 $\mu$m, its length of 1 cm and 
%a target volume height of 100 $\mu$m 
%we get the target volume: $4 \ 10^{-5} \ \mathrm{cm^3}$. 

An estimate of the target density is obtained by assuming all Ps atoms
to fly perpendicularly from the crystal surface~\cite{Ps_emission}. 
The number of positrons stored in the plasma is $N_{e^+} \ = \ 10^{11}$. 
The positron flux from the plasma to the crystal has a duration 
of $\sim$ 10 ns. 
The Ps atoms spend only 0.100 ns in the target volume, hence the 
target density is given by:

\begin{eqnarray}
n_{P_s} \ = \  \frac{10^{11}}{10^{-4}}\frac{0.100}{10} \ \epsilon_{P_s} \ \epsilon_{CFAC}= 
\  0.25 \ 10^{13} \ \mathrm{cm^{-3}}
\label{eq:targ-dens}
\end{eqnarray}

The target volume could also be counted starting from the crystal surface with some 
antiproton losses (antiprotons hitting the crystal). 

The simulation of the CFAC gives an efficiency 
of $\epsilon_{CFAC}$ = 0.5 for a positron to reach the crystal.
The positronium emission efficiency from the crystal surface is taken
as $\epsilon_{P_s}=0.5$.

A correcting factor due to the positronium decay between the crystal 
and the target volume will reduce the density by less than a factor
2. This correction depends on the magnetic field strength.

\subsection{Positron losses}
When mixing $e^{-}$ and $e^{+}$ with a density above 
$\sim 10^9 \ \mathrm{cm^{-3}}$
and a temperature of a few meV, the plasma collapses into positronium
through a 3 body interaction. 

\begin{eqnarray}
e^{-} +  e^{+} + e^{\pm} \ & \longrightarrow  & 
\ \mathrm{P_{S}(n'l')} + e^{\pm} 
\label{eq:eee}
\end{eqnarray}

The rate of other reactions, 
namely the radiative recombination and the direct annihilation are negligible. 
This reaction and the subsequent positronium annihilation destroy the 
positrons. 

The number of positrons in the neutral plasma as a function of time is
given by~\footnote{see computation in appendix}:
\begin{eqnarray}
N_e = \sqrt{\frac{a}{\lambda}} \times \tanh(\sqrt{a \ \lambda} \ t) \\
\lim_{t \to \infty}N_e(t) = \sqrt{\frac{a}{\lambda}}
\label{eq:dNedt0}
\end{eqnarray}

Where $a$ is the injection rate and $\lambda$ is a constant related to 
the 3 body reaction.

The tables obtained in the appendix show that in the worst case, 
in $\sim$ 2 s and at a 
temperature equal or above 4 meV it is possible to accumulate 
$2 \ 10^{11} \ e^{+}$ in the plasma. For less than 10 seconds
of accumulation there is little variation with temperature above 4 meV.  
Once the injection of positrons ends and during $\sim$ 1s 
the loss of positrons is small and on the time scale of a few $\mu$s 
it can be neglected. 

In order to have a negligible loss by annihilation on remaining gas a
vacuum environment is required with a residual gas partial pressure 
below $\sim 10^{-9}$ torr.

\section{Antihydrogen production}
\label{antihprod}

\par 
The charge exchange on positronium to produce the atom is a resonant reaction. 
The cross-section calculations~\cite{Ps-catalysis4,Ps-catalysis1,Ps-catalysis2} 
have been confirmed by an experiment~\cite{Ps-catalysis3,Ps-p-experiment}.

\par
Once the $\mathrm{\overline{H}}$ atom is produced, it captures a positron to
make the ion. There is a non resonant 
channel, the two body recombination, and a resonant one, the charge 
exchange on positronium.

\par
\begin{figure}
%\vspace{-1cm}
\begin{center}
%\begin{picture}(100,90)
%\put(10,10){\epsfxsize85mm\epsfbox{fig-soudure1.eps}}
\includegraphics*[width=85mm]{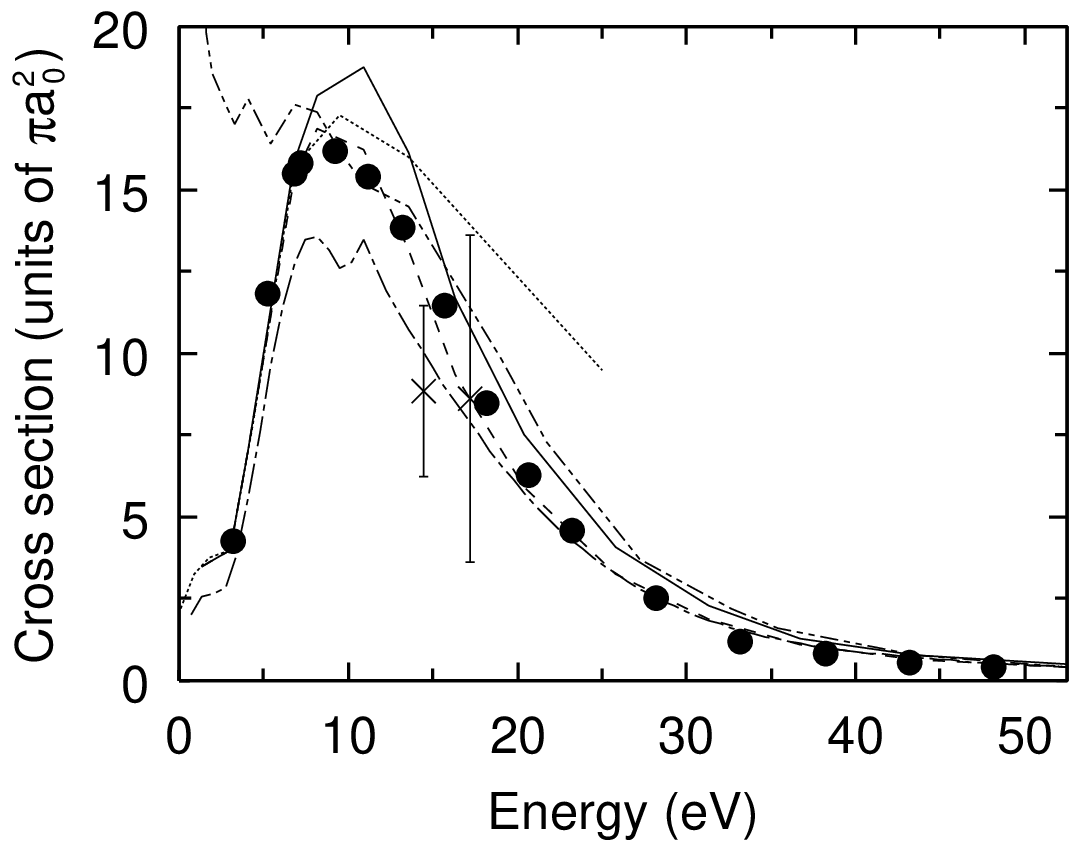}
%\end{picture}
\end{center}
\vspace*{-0.3cm} 
\caption{\protect\footnotesize 
Hydrogen formation cross section in 
$\mathrm{P_{S} + p \rightarrow H} + e^+$ collisions~\cite{Ps-catalysis4}.}
Closed circles represent the calculation of Yamanaka and
Kino; lines, other models referenced in
the same paper; 
crosses, the experiment of Merrison et al.~\cite{Ps-catalysis3,Ps-p-experiment}.
The abscissa is the energy of the proton in the center of mass frame.
\label{fig:yamanaka}
\end{figure}

\begin{figure}
%\vspace{-1cm}
\begin{center}
%\begin{picture}(100,90)
%\put(10,10){\epsfxsize85mm\epsfbox{fig-soudure1.eps}}
\includegraphics*[width=85mm]{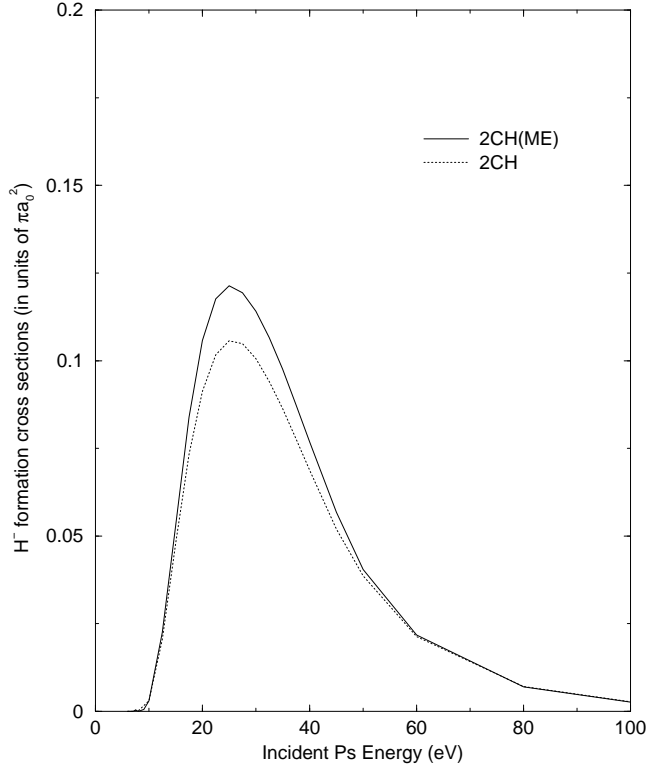}
%\end{picture}
\end{center}
\vspace*{-0.4cm} 
\caption{\protect\footnotesize 
$\mathrm{H^-}$ production cross section in 
$\mathrm{H}(n=1) + \mathrm{P_{S}}(n'=1) \rightarrow \mathrm{H^-} + e^+ $ 
collisions~\cite{HPs}.
The lines are two variants of a two-channel calculation by
Biswas.
The abscissa is the energy of the Ps in the center of mass frame.
}
\label{fig:biswas}
\end{figure}

 The choice of the antiproton kinetic energy is given by the overlap of 
the resonant cross sections. In the region of interest, the
antiprotons have a kinetic energy of $\sim$ 13 keV 
(figure~\ref{fig:yamanaka}). 
The atom production cross section obtained 
experimentally~\cite{Ps-catalysis3} is 
7.8 $\pi a_{0}^{2}$~\footnote{Here $a_0$ is the Bohr radius and 
$\pi a_{0}^{2} = 0.880 \times 10^{-16} \ \mathrm{cm^2}$.}. 
At this kinetic energy, the computed cross section~\cite{HPs} for the ion 
production is $\sim$ 0.05 $\pi a_{0}^{2}$ (figure~\ref{fig:biswas}).

\par
The (anti)hydrogen production rate normalized to the (anti)proton flux is:
\begin{eqnarray}
n_{P_{s}} \times \sigma_{\mathrm{\overline{H}}} \times v_{\overline{p}}
\label{eq:hbar-rate}
\end{eqnarray}

The (anti)proton speed at 13 keV is $v_{P} = 1.58 \ 10^{8} \mathrm{cm s^{-1}}$.
Therefore the normalized rate is 0.28 $10^6 \ s^{-1}$.
The target length is 1 cm, 
hence the crossing time by the (anti)proton beam is 6.3 ns and the number of
antihydrogen atoms produced by each antiproton crossing the target is 0.0017.

\par
The (anti)hydrogen ion production rate normalized to the antihydrogen flux is:
\begin{eqnarray}
n_{P_{s}} \times \sigma_{\mathrm{\overline{H}}^+} \times 
v_{\mathrm{\overline{H}}}
\label{eq:hbar-rate2}
\end{eqnarray}

The normalized rate is 0.18 $10^4 \ s^{-1}$. After convolution for 
the $\overline{H}$ production in the same 1 cm long target, 
the number of ions produced for each antiproton crossing the 
target is $10^{-7}$.

At the edge of the crystal and at a distance of 150 $\mu$m the CFAC 
electrostatic field component transverse to the antiproton path is 
$\sim$ 0.12 Mega Volts/m. Therefore an antiproton aiming at 
the edge of the crystal will be pulled transversally by 240 $\mu$m 
on a distance of 1 cm.
The transversal electrostatic field which is null on the axis and 
grows toward the edges of the target,
will cause a maximum pull $\sim$ 0.5 mm and a beam divergence of 
$\sim$ 50 mrd.

\section{Discussion}
\label{discuss}
\par
In the integration of the devices (CFAC, MCEO trap, buffer-gas, 
Greaves-Surko trap) into a single experiment there is room
for changing their parameters:
%. But we have avoided making changes:
gains from tuning the device parameters
and losses are difficult to simulate and have to be established 
experimentally. When changing the parameters in the simulation these
gains/losses are within a factor 10. 

Still there are many technical questions which are not addressed
by the simulation. We shall now list some of them.

The field in the buffer gas trap is 1.5 kGauss and 5 T in the 
Greaves-Surko trap:
a set of iron plates and auxilliary fields shall be designed to
reach the 100 Gauss guiding solenoidal field~\cite{traplim}. 

Positron plasmas with a density of $10^{10} e^{+} \mathrm{cm^{-3}}$ 
at meV temperatures 
have been stored in a Greaves-Surko trap with a few 
$10^{8}$ trapped $e^{+}$. 
The next generation of multicell traps~\cite{traplim,traps,surko-trap}
is expected to store and cool to meV temperature $10^{13} e^{+}$. 
The fast loading mode
cannot be implemented without traps that can store at least $10^{11} e^{+}$.
In this fast mode, once trapped, the positrons shall be extracted in 
a single pulse. This step has to be experimentally established.

Such traps cannot 
be filled with $e^{+}$ emitted by a radioactive source in less than a few 
weeks. An intense source of slow positrons such as the one we 
proposed~\cite{perez-roso}, using an intense 10 MeV electron beam
impinging at a small incidence angle on a thin tungsten foil
is expected to fill the trap at a rate of $10^{10} e^{+}
\mathrm{s^{-1}}$.

In the slow loading mode, no trap is needed and only a buffer gas is required.

At high density, the neutral plasma is not transparent to the electro-magnetic field.
The extraction of the positrons by the CFAC field in $\sim $ 10 ns requires
a pulse shaping technique~\cite{chen} and shall be tested experimentally.

In order to implement a gravitation experiment with a scheme such as
the one discussed by J. Walz and T. W. H\"{a}nsch~\cite{walz}
constraints have to be met by the antiproton source.
The ions are produced in a beam at 13 KeV with an energy and angular spread 
dominated by the initial (anti)proton energy spread and the 
$\sim$ 50 mrd divergence due to the CFAC field. The angular 
spread due to the charge exchange reactions is negligible. The angular spread
shall be taken into account to decelerate and collect the 
$\mathrm{\overline{H}^+}$ into an ion trap with high 
efficiency~\cite{toshiyasu}. Antiproton traps already exist, 
but a trap with $10^6  \mathrm{\overline{p}}$ 
capable of producing a beam of 13 keV $\pm$ 1 keV energy with a 
divergence lower than 0.01 rd (ie 100 $\mu$m on 1 cm path) is still 
to be demonstrated.

\par
For antihydrogen creation, the very large $e^{\pm}$ plasma densities
discussed above allow to transform the antiprotons with a relatively
high efficiency. But antiprotons are difficult to produce. 
However, in order to test the process one can use
protons and an existing trap: the lower plasma density
being compensated by a larger number of protons (say $10^{8}$ p  
per beam pulse). 

\par
Another aspect of this path for a gravity experiment is that the reactions 
involved are at all steps charge symmetric: by switching from protons to 
antiprotons on the time scale of 1 hour, it is possible to measure $g$
in the same gravitational field with little or no tide effect due the 
movement of the moon and other masses.

\par
\begin{figure}
\vspace{-1cm}
\begin{center}
%\begin{picture}(100,90)
%\put(10,10){\epsfxsize85mm\epsfbox{fig-soudure1.eps}}
\includegraphics*[width=90mm]{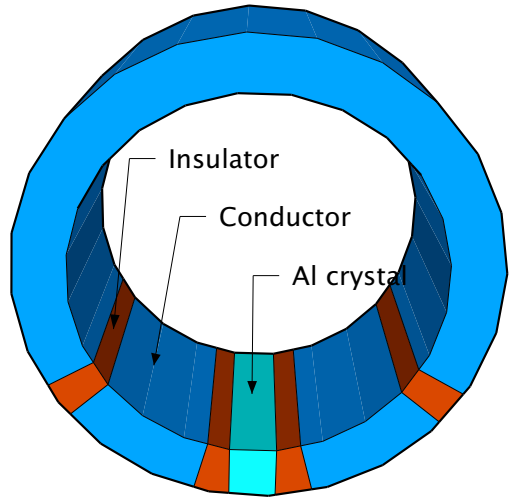}
%\end{picture}
\end{center}
\vspace*{-2.5cm} 
\caption{\protect\footnotesize 
Modified central electrode of a MCEO trap to produce an approximate 
CFAC field.}
\label{fig:musashi-msgc}
\end{figure}

The integration of the CFAC into a MCEO trap requires some
modification.
We foresee to cut the central electrode in 4 (or more) sections separated 
by a thin insulator (figure~\ref{fig:musashi-msgc}).
In the trapping mode all the sections are set at the same potential 
acting as a single electrode.
In the CFAC mode the potentials are set to produce an electrical field 
similar to the one in figure ~\ref{fig:msgc1a}.

\par
The $\mathrm{\overline{H}^+}$ ion beam is needed for the gravity
experiment, but a large number of
$\mathrm{\overline{H}}$ atoms are also produced in the beam. Using a 
laser to stimulate the transition toward the $n=2$ level one gets a
beam in the $\sim$10-20 keV range.
Then the Separated Oscillatory Field method~\cite{SOF-1,SOF-2,SOF-3} 
allows to measure the $2S_{1/2}-2P_{3/2}$ fine structure.

{\bf Appendix}

\par
The technique described in this article can be used for experiments 
on positronium BEC and a 511 keV laser because it could reach a 
positronium density of $\sim 10^{14}$ or $\sim 10^{15}$ cm$^{-3}$.
The appendix gives a numerical estimation of the density
as a function of time and temperature.

\par
The maximum number of positrons available to interact with an
Aluminium crystal is the result of the equilibrium between the
injection rate and the 3 body reaction rate.
This reaction (see equation~\ref{eq:eee}) and the subsequent 
positronium annihilation destroy the positrons. 

The time evolution of the number of positrons in the plasma is given
by the injection rate and the 3 body reaction rate~\cite{glinsky}:
$dN_e = a \ dt - r_3 \ dt$ 

where $N_e$ is the number of $e^+$ in the plasma volume, 
$a$ is the injection rate and
$r_3$ is the 3 body reaction rate given by:
$r_3 = A \times n_{e}^2 \times v_{e} \times b^5$

where  the electron thermal velocity $v_{e} = \sqrt{\frac{kT}{m_{e}}}$
and the impact parameter $b = \frac{e^{2}}{kT}$ are numerically:
$v_e (\mathrm{cm.s^{-1}}) = 4.19 \ 10^7 \sqrt{T(\mathrm{eV})}$ and 
$b (\mathrm{cm.eV^{-1}}) = 1.44 \ 10^{-7} / T(\mathrm{eV})$, resulting in
$r_3 (\mathrm{s^{-1}}) = 2.6 \ 10^{-27} \ A \times n_{e}^{2} \ / \ T^{4.5}$ 

where $A$ is a parameter which varies with the magnetic field, 
$n_e$ is the density in $\mathrm{cm^{-3}}$ and $T$ is the plasma 
temperature in eV. 

The neutral plasma volume is $\sim 30 \ \mathrm{mm^3}$. 
 
Let $V$ be the plasma volume and let's define
$\lambda$ as $\lambda = 2.6 \ 10^{-27} \ A  / (V^{2} \ T^{4.5}) $,
so that $r_3 = \lambda \ N_{e}^2$. 

With the initial condition $N_e (t=0) =
0$, we get:
$dN_e = (a - \lambda \ N_e^{2}) \ dt$

Which leads to:
$N_e = \sqrt{\frac{a}{\lambda}} \times \tanh(\sqrt{a \ \lambda} \ t)$ , $ 
\lim_{t \to \infty}N_e(t) = \sqrt{\frac{a}{\lambda}}$

For the two extreme situations, $B=0$ and $B= \infty$, the $A$
parameter is respectively equal to 0.76 and 0.07.

In the computations the ion was supposed to be much heavier than the electron
and therefore its trajectory in the field was neglected~\cite{glinsky}. 
Recently a computation was made with a proton taking into account 
its trajectory~\cite{robicheaux} for fields of a few Tesla: 
the variation of the parameter $A$ compared to the infinite field
value was less than a factor 2. 
Here the ion is the positron which is much lighter and the magnetic 
field is a few Gauss only. Therefore the positron behaves as a 
heavier ion in a stronger field.
The expected behavior is then an intermediate one, between the above 
computation with 0 field and with infinite field.   

A computation involving only electrons and positrons predicts a three
body recombination rate twice greater than for protons and 
electrons~\cite{Hoepker}.

The computed number of positrons accumulated inside the neutral 
plasma for several temperatures is tabulated for both values of
$A$.

\begin{table}
\centering
\begin{tabular}{|c|cccccc|}
\hline
T (meV)               &   2   &   4   &   10      &   25     &    50     &  100    \\
\hline

0.1 s & 9.99 $10^{9}$ & 9.99 $10^{9}$ & 9.99 $10^{9}$ & 9.99 $10^{9}$ & 9.99 $10^{9}$ & 1.0 $10^{10}$ \\

1 s   & 9.09 $10^{10}$ & 9.95 $10^{10}$ & 9.99 $10^{10}$ & 9.99 $10^{10}$ & 9.99 $10^{10}$ & 1.0 $10^{11}$ \\

2 s   & 1.45 $10^{11}$ & 1.96 $10^{11}$ & 1.99 $10^{11}$ & 1.99 $10^{11}$ & 1.99 $10^{11}$ & 2.0 $10^{11}$ \\

\hline
\end{tabular}
\caption{Number of $e^{+}$ accumulated in the 
$30 \ \mathrm{mm^{3}}$ neutral plasma with an injection rate $a = 10^{11}s^{-1}$ for A = 0.76 (B = 0)}
\label{TAB:1}
\end{table}

\begin{table}
\centering
\begin{tabular}{|c|cccccc|}
\hline
T (meV)               &   2   &   4   &   10      &   25     &    50     &  100    \\
\hline

0.1 s & 9.99 $10^{9}$ & 9.99 $10^{9}$ & 9.99 $10^{9}$ & 9.99 $10^{9}$ & 1.0 $10^{10}$ & 1.0 $10^{10}$ \\

1 s   & 9.91 $10^{10}$ & 1.0 $10^{11}$ & 1.0 $10^{11}$ & 1.0 $10^{11}$ & 1.0 $10^{11}$ & 1.0 $10^{11}$ \\

2 s   & 1.93 $10^{11}$ & 2.0 $10^{11} $ & 2.0 $10^{11} $ & 2.0 $10^{11} $ & 2.0 $10^{11} $ & 2.0 $10^{11} $ \\

\hline
\end{tabular}
\caption{Number of $e^{+}$, accumulated in the  
$30 \ \mathrm{mm^{3}}$ neutral plasma with an injection rate $a = 10^{11}s^{-1}$ for A = 0.07 (B = $\infty$)}
\label{TAB:2}
\end{table}

\begin{table}
\centering
\begin{tabular}{|c|cccccc|}
\hline
T (meV)     &   2   &   4   &   10      &   25     &    50     &  100    \\
\hline

A = 0.76 & 1.81 $10^{11}$ & 8.60 $10^{11}$ & 6.76 $10^{12}$ & 5.31 $10^{13}$ & 2.53 $10^{14}$ & 1.20 $10^{15}$ \\

A = 0.07 & 5.95 $10^{11}$ & 2.83 $10^{12}$ & 2.23 $10^{13}$ & 1.75 $10^{14}$ & 8.32 $10^{14}$ & 3.96 $10^{15}$ \\

\hline
\end{tabular}
\caption{$\sqrt{\frac{a}{\lambda}} $ = maximum number of $e^{+}$, 
that can be accumulated in the  
$30 \ \mathrm{mm^{3}}$ neutral plasma with an injection rate $a = 10^{11}s^{-1}$ for A = 0.76 (B = 0) and 
A = 0.07 (B = $\infty$)}
\label{TAB:3}
\end{table}

When the injection in the plasma ends, the evolution is given by the 
rate of the 3 body reaction:

%\begin{eqnarray}
$dN_e = - r_3 \  dt = - \lambda  \ N_e^{2} \ dt \\
%N_e(t=0) = N_{0} \\
N_e = \frac{N_{0}}{1 + N_{0} \ \lambda \ t} $
%\label{eq:Ne-down}
%\end{eqnarray}

{\bf Acknowledgements}\\
We wish to express our sincere thanks to all the people from different 
fields with whom we had fruitful discussions:
P.K.Biswas,
G. Chardin,
C. Guyot,
B. Mansouli\'{e},
C. Surko,
R. Greaves,
A.P. Mills,
J.M. Rax,
Y. Yamazaki.\\
One of us is grateful to Chi-Yu Hu~\cite{Ps-catalysis1} who carefully 
presented the use of
positronium to produce antihydrogen at the poster session of the ICPEAC 2001 
conference.

\end{document}